\documentclass[aps,pra,twocolumn,superscriptaddress,10pt,showpacs]{revtex4-1}

\pdfoutput=1

\usepackage{amsmath,amsthm,amssymb} %
\usepackage[english]{babel} %
\usepackage[singlelinecheck=off,justification=raggedright]{caption} %
\usepackage{color} %
\usepackage{csquotes} %
\usepackage{dsfont} %
\usepackage{times} % charter, fourier, mathpazo, times
\usepackage{graphicx} %
\usepackage{subfig}
\usepackage[breaklinks]{hyperref} %
\usepackage{suffix} % for *-version commands
\usepackage{xspace} %

\hypersetup{colorlinks=true, linkcolor=blue, citecolor=magenta}

\newcommand{\ket}[1]{\ensuremath{|#1\rangle}} %
\newcommand{\real}{\mathbb{R}} %
\DeclareMathOperator{\tr}{tr} %
\begin{document}

%% End-Of-Header

\title{Geometry of reduced density matrices for symmetry-protected
  topological phases}

\author{Ji-Yao Chen} %
\affiliation{State Key Laboratory of Low Dimensional Quantum Physics,
  Department of Physics, Tsinghua University, Beijing, China} %
\affiliation{Perimeter Institute for Theoretical Physics, Waterloo,
  Ontario, Canada} %
\author{Zhengfeng Ji} %
\affiliation{Institute for Quantum Computing, University of Waterloo,
  Waterloo, Ontario, Canada} %
\affiliation{State Key Laboratory of Computer Science, Institute of
  Software, Chinese Academy of Sciences, Beijing, China} %
\author{Zheng-Xin Liu} %
\affiliation{Institute for Advanced Study, Tsinghua University,
  Beijing, China} %
\author{Yi Shen} %
\affiliation{Department of Statistics and Actuarial Science,
  University of Waterloo, Waterloo, Ontario, Canada} %
\author{Bei Zeng} %
\affiliation{Department of Mathematics \& Statistics, University of
  Guelph, Guelph, Ontario, Canada} %
\affiliation{Institute for Quantum Computing, University of Waterloo,
  Waterloo, Ontario, Canada} %

\date{\today}
\begin{abstract}
  In this paper, we study the geometry of reduced density matrices for
  states with symmetry-protected topological (SPT) order. We observe
  ruled surface structures on the boundary of the convex set of low
  dimension projections of the reduced density matrices. In order to
  signal the SPT order using ruled surfaces, it is important that we
  add a symmetry-breaking term to the boundary of the system---no
  ruled surface emerges in systems without boundary or when we add a
  symmetry-breaking term representing a thermodynamic quantity.
  Although the ruled surfaces only appear in the thermodynamic limit
  where the ground-state degeneracy is exact, we analyze the precision
  of our numerical algorithm and show that a finite system calculation
  suffices to reveal the ruled surface structures.
\end{abstract}

\pacs{03.67.-a, 03.65.Ud, 03.67.Dd, 03.67.Mn}

\maketitle

\section{Introduction}

The idea of reading physical properties from the geometry of the
underlying convex bodies arising naturally from quantum many-body
physics has been examined and re-examined many times from different
perspectives in the literature~\cite{Col63,Erd72,EJ00,VC06,GM06,SM09}.
It appeared in the context of the $N$-representability problem in
quantum chemistry and, more generally, the quantum marginal problem in
quantum information theory~\cite{Col63,klyachko2006quantum}. Recently, the approach
received revived interests and was used in the investigations of
quantum phases from the convex geometry of reduced density matrices or
their low dimension
projections~\cite{EJ00,VC06,GM06,SM09,chen2014principle}.

Consider a many-body local Hamiltonian $H = \sum_j H_j$ where each
term $H_j$ acts non-trivially on a set of particles $S_j$. Usually,
each $S_j$ contains only a constant number of neighboring particles
and defines the geometry of the local interactions of the system. The
convex set (a convex set in Euclidean space is the region that, for every pair of points within the region, every
point on the straight line segment that joins the pair of points is also within the region. For instance, a solid sphere 
forms a convex set) of quantum marginals, the collections of reduced density
matrices $(\rho_j)$, is of fundamental importance to quantum many-body
physics. Here, each $\rho_j$ in the same tuple is the reduced state on
$S_j$ of a common many-body state. The convex set of quantum marginals
is independent of the particular form of the Hamiltonian $H$, but
depends only on the geometric locality of the system defined by
$S_j$'s. Should we have a complete characterization of the quantum
marginals, most problems of many-body physics would become extremely
easy. As expected, however, the structure of the set of quantum
marginals is rather complicated.

To study quantum phase transitions, it seems appropriate to
investigate a further coarse-grained set of the quantum marginals. The
usual situation one often faces when considering quantum phase
transitions at zero temperature is the following: the Hamiltonian of
the system has the form $H = J_1 H_1 + J_2 H_2$ and one is concerned
with the change of the properties of lower energy states of $H$ as
$J_1$ and $J_2$ change. The terms $H_1$ and $H_2$ now act on a large
number of particles, but they are still sums of local terms. Let the
convex set $\Theta(\{H_j\}) \subseteq \real^2$ be the set of all
points $(\tr(H_1\rho),\tr(H_2\rho))$ for $\rho$ ranging in the set of
all possible many-body states. This set is a low dimension projection
of the convex set of quantum marginals and is hopefully much easier to
analyze.

It is obvious that for any $(\alpha_1,\alpha_2)$ in $\Theta(\{H_j\})$,
$\sum_i J_i \alpha_i \ge E_0(H)$, the ground state energy of $H$. This
means that the Hamiltonian $H$ can be thought of as the supporting
hyperplane of $\Theta(\{H_j\})$, and the change of parameters
$J_1, J_2$ can be visualized as the change of the supporting
hyperplane, moving around the convex set. The intersection of this
hyperplane with $\Theta(\{H_j\})$ corresponds to the image of the
ground state of $H$ on the boundary of $\Theta(\{H_j\})$. A flat
portion on the boundary of $\Theta(\{H_j\})$ signatures the
first-order phase transition. However, for continuous phase
transitions, the geometry of $\Theta(\{H_j\})$ alone does not convey
any informative signals~\cite{chen2014principle}.

Recently, the convex geometry approach was employed in the study of
quantum symmetry-breaking phases~\cite{zauner2014symmetry}. For a
Hamiltonian $H = J_1 H_1+ J_2 H_2$ with certain symmetry, one adds a
third, symmetry-breaking, term $H_3$ to the Hamiltonian and consider
$H = J_1 H_1 + J_2 H_2 + J_3 H_3$. The authors plotted the convex set
$\Theta(\{H_j\}) \subseteq \real^3$ of points
$(\tr(H_1\rho), \tr(H_2\rho), \tr(H_3\rho))$ and analyzed the geometry
of its boundary. On this set, the emergence of ruled surfaces on the
boundary is observed (a ruled surface is a surface that can be swept
out by moving a line in space, or equivalently, for any point on the ruled surface
there exists a line passing through this point that is also on the surface. For example, the 
curved boundary of a cylinder is a ruled surface). The authors of~\cite{zauner2014symmetry} argued that the existence of
those ruled surfaces is a defining property of symmetry-breaking and
can be used to signal symmetry-breaking phase transition.

Interestingly, the observation that ruled surfaces on the boundary of
certain convex body can explain phase transitions dates back to Gibbs
in the 1870's~\cite{Gib73a,Gib73b,Gib75,Isr79}, even though the convex
bodies under consideration in classical thermodynamics and quantum
many-body physics are rather different. It indicates that the convex
geometry approach is a rather fundamental and universal idea.

In the present paper, we study the phenomenon of the emergence of
ruled surface on the boundary of the convex set
$\Theta(\{H_j\}) \subseteq \mathbb{R}^3$ for one dimensional (1D) symmetry-protected
topological (SPT) ordered systems. An SPT ordered state is a
bulk-gapped short-range entangled state with symmetry protected
nontrivial boundary excitations~\cite{GW0931}. The well known two dimensional  and
three dimensional topological
insulators~\cite{KM0501,BZ0602,KM0502,MB0706,FKM0703,QHZ0824} are free
fermion SPT phase protected by time reversal and $U(1)$ charge
conservation symmetry, whose boundary remain gapless as long as the
symmetries are preserved. SPT phases also exist in interacting
systems. A typical example of interacting bosonic SPT phase is the 1D
spin-1 Haldane chain~\cite{H8350,H8364,AKL8877,GW0931,PBT0959}, whose
degenerate edge states are protected by time reversal, or spatial
inversion, or $Z_2\times Z_2$ spin rotation
symmetry~\cite{GW0931,PBT0959}. Bosonic SPT phases can be partially
classified by group cohomology theory~\cite{CGL1314, CGLW2012}.
Especially, in 1D (which we will study in this paper), SPT phases with
onsite-symmetry are classified by $\mathcal H^2(G, U(1))$, or the
projective representations of the symmetry group
$G$~\cite{ChenGuWen2011_1D, ChenGuWen2011_1Dfull, Cirac2011}.

The ground-state degeneracy is a necessary condition for
the existence of a ruled surface. However, in order
to observe such a ruled surface, we need to add a local term that can lift the
degeneracy. In other words, ground-state degeneracy
that can be lifted by some local term will lead to ruled surfaces.
It is then required such a
local term exists. In case of the symmetry-breaking order, it corresponds to
the local order parameter. We show that in case of the SPT order, the
emergence of a ruled surface only exists for system under open boundary condition (OBC),
with the corresponding local term acting on the boundary that breaks
the symmetry of the system, hence lifting the ground-state degeneracy.

To show this, we study a 1D model exhibiting $Z_2\times Z_2$ SPT
order. We discuss in detail the effect of geometric locality and its
relationship to the emergence of ruled surfaces. Since the degeneracy
of the ground states is only exact in thermodynamic limit, in
principle the ruled surface also requires such a limit. However,
numerical results suggest that in practice, the ruled surface can already be observed
for large but finite systems. This allows us to study the features of ruled surface
based on finite-system calculations.

One important difference between our results and
Ref.~\cite{zauner2014symmetry} is that the boundary terms that lift
the degeneracy are not associated with thermodynamic variables. It is
essentially the effect of geometric locality (i.e. boundary conditions
do change the geometric locality of the system) that leads to a
different geometry of the set of reduced density matrices. On the
contrary, for a topological ordered system, no local terms can lift
the topological degeneracy; therefore one cannot observe ruled surface
on the geometry of local reduced density matrices. Our results hence lead to a
deeper understanding of the physical meaning of the emergence of ruled
surface.

% todo: We organize our paper as follows.

\section{$Z_2\times Z_2$ SPT order: The 1D cluster state in magnetic
  fields}

\begin{figure}
\centering
\subfloat[]{
\includegraphics[width = 85mm,height = 9mm]{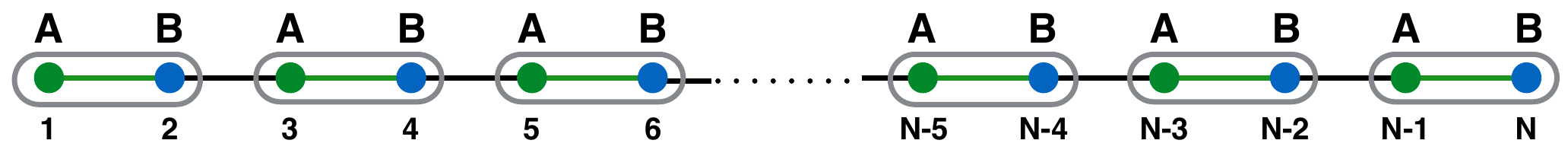}}\\
\subfloat[]{
\includegraphics[width=55mm,height=55mm]{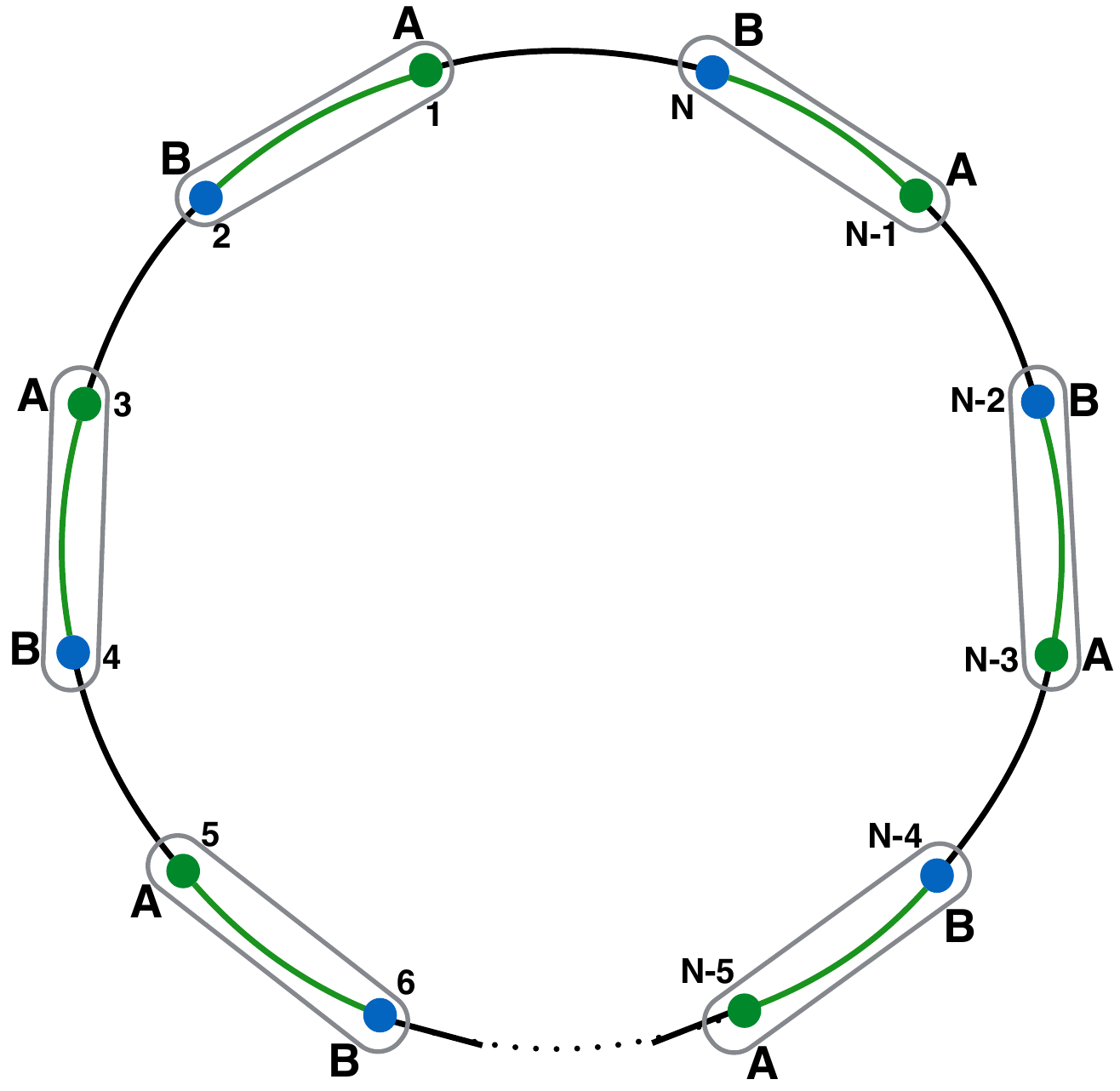}}
\caption{1D system with A, B sublattice. The number of sites $N$ is
    even. Green and blue dots represent sublattice A and B,
    respectively. Each small circle represents a unit cell. (a) is the 1D system under open
    boundary condition and (b) is the system under periodic boundary condition.}
\label{1D system}
\end{figure}

Consider a 1D system with A, B sublattice under OBC, see Fig.~\ref{1D system}(a). Assuming the number of sites $N$
is even, the Hamiltonian reads:

\begin{equation}
  H_{clu}(B_z) =\sum_{i=2}^{N-1} X_{i-1}Z_{i}X_{i+1} + B_z\sum_{i=1}^{i=N}Z_i,
\end{equation}
where $X, Z$ are Pauli operators.

The first term $X_{i-1}Z_{i}X_{i+1}$ corresponds to the stabilizer
generators of the $1D$ cluster state (without boundary
terms)~\cite{briegel2001persistent,raussendorf2001one}. The second
term corresponds to a longitudinal magnetic field. Notice that $H_{clu}$
has a $Z_2\times Z_2$ symmetry generated by the following two
operators (see
e.g.~\cite{else2012symmetry,skrovseth2009phase,doherty2009identifying}),
\begin{align}
  O_1 &= Z_1Z_3Z_5...Z_{N-1}\\
  O_2 &= Z_2Z_4Z_6...Z_N
\end{align}
Obviously, $[O_1, O_2]=0$ and $[O_1,H_{clu}] = 0, [O_2,H_{clu}]=0$. In
fact, there is also a hidden continuous $U(1)$ symmetry in $H_{clu}$
generated by $\sum_{i=1,odd}^{i=N-1} Y_iX_{i+1} - X_iY_{i+1}$.

When $B_z<1$, the ground state of above model has
  $Z_2\times Z_2$ SPT order. To show this, we can transform
$H_{clu}$ into a familiar form with only two-body
interactions~\cite{zeng2014topological}. Consider unitary operations
$U_{AB}$ acting on each nearest A-B sites, as denoted in Fig.\ref{1D system}. Each
$U_{AB}$ transform Pauli operators as follows:
\begin{align}
  X_A &\to X_A, & Z_A&\to Z_AX_B,\\
  X_B &\to X_B, & Z_B&\to X_AZ_B.
\end{align}
In fact, $U_{AB}$ is nothing but the controlled-$Z$ operation (i.e.
$CZ=\text{diag}(1,1,1,-1)$) in the Pauli $X$ basis (i.e.
$\frac{1}{\sqrt{2}}(\ket{0}\pm\ket{1})$). That is,
$U_{AB}=(H_A\otimes H_B)CZ_{AB}(H_A\otimes H_B)$, where $H$ is the
Hadamard transformation given by $HXH=Z$.

Under this transformation, we have
\begin{align}
  X_{i-1}^BZ_i^AX_{i+1}^B\to X_{i-1}^BZ_i^A\\
  X_{i-2}^AZ_{i-1}^BX_i^A\to Z_{i-1}^BX_i^A
\end{align}
and
\begin{align}
  Z_i^A\to Z_i^AX_{i+1}^B\\
  Z_{i+1}^B\to X_i^AZ_{i+1}^B.
\end{align}

Thus we can recast the original Hamiltonian as following:
\begin{equation}
  \label{eq:HOBC}
  \begin{split}
    H_{OBC} &= J_1(1+\alpha)
    \sum_{i=2,even}^{N-2} (X_iZ_{i+1} + Z_iX_{i+1}) \\
    &+ J_2(1-\alpha) \sum_{i=1,odd}^{N-1}(X_iZ_{i+1} +
    Z_iX_{i+1}).\\
  \end{split}
\end{equation}
For the convenience of later calculation, we let $J_1=\pm1$, $J_2=\pm1$, $-1 \leq \alpha \leq 1$.
 If we further apply an unitary transformation on the A sublattice
such that
\[
X_i^A\to Z_i^A,\ \ Z_i^A\to X_i^A,
\]
then the model (\ref{eq:HOBC}) becomes the familiar XY model~\cite{Perk1,Perk2,Perk3,Perk4},
\begin{equation}
  \begin{split}
    H_{OBC} &= J_1(1+\alpha)\sum_{i=2,even}^{N-2} (X_iX_{i+1} + Z_iZ_{i+1}) \\
    &+ J_2(1-\alpha)\sum_{i=1,odd}^{N-1}(X_iX_{i+1} + Z_iZ_{i+1}).
  \end{split}
  \label{XY}
\end{equation}

The $Z_2\times Z_2$ symmetry becomes very clear in above
  Hamiltonian: it is generated by the uniform $X$ and $Z$ operations.
  Since in the above model each unit cell contains two spins,
  the strong bonds may locate inter unit cells or intra unit cells,
   and consequently there are two different phases.
  When $0 < \alpha\leq1$, the ground state carries nontrivial $Z_2\times Z_2$
  SPT order and has two-fold degenerate edge states (which carry projective representation $\{I,X,Y,Z\}$ of $Z_2\times Z_2$ group) on each boundary;
  on the contrary, when $-1\leq\alpha<0$, the model falls in a trivial
  symmetric phase without edge states; $\alpha=0$ corresponds to the phase transition point.
  The previously mentioned $U(1)$
  continuous symmetry is generated by $\sum_i Y_i$ in Eq. (\ref{XY}).
  The $U(1)$ symmetry is an accidental symmetry for the
  SPT order, namely, the properties of the two phase remains unchanged
  if the $U(1)$ symmetry is destroyed by anisotropic interactions.

  In later discussion, we will go back to the original cluster model. For the
  purpose of studying the reduced density matrix, we further add a
  transverse magnetic field. We will consider cases with both OBC and
  periodic boundary condition (PBC) .
  Thus under OBC (see  Fig.~\ref{1D system}(a)), the Hamiltonian reads:
  \begin{equation}
  \label{H_OBC}
  \begin{split}
    H_{OBC}
    &= J_1(1+\alpha)\sum_{i=2}^{N-1} X_{i-1}Z_{i}X_{i+1}\\
    &+J_2 (1-\alpha)\sum_{i=1}^{N}Z_i -
        B_x\sum_{i=1}^{N}X_i\\
    &= J_1(1+\alpha)
        H^O_1+J_2(1-\alpha)
        H^O_2-B_xH_3,\\
  \end{split}
  \end{equation}
  while under PBC (see Fig.~\ref{1D system}(b)), the Hamiltonian is:
  \begin{equation}
  \begin{split}
    H_{PBC}
    &= J_1(1+\alpha)\sum_{i=2}^{N+1} X_{i-1}Z_{i}X_{i+1}\\
    &+ J_2 (1-\alpha)\sum_{i=1}^{N}Z_i -
        B_x\sum_{i=1}^{N}X_i \\
    &= J_1(1+\alpha)
        H^P_1+J_2(1-\alpha)
        H^P_2-B_xH_3,\\
  \end{split}
  \end{equation}
  where the $(N+1,N+2)$ sites are identified with  the $(1,2)$ sites, respectively.

\section{Effect of locality and the emergence of ruled surface for SPT
  phase}

One necessary condition for the emergence of a ruled surface is the
ground-state degeneracy. When $B_x=0, (J_1,J_2)=(\pm 1,\pm 1), 0<\alpha\leq1$, the ground-state of $H_{OBC}$
is four-fold degenerate (if $N\to\infty$) and the ground state of
$H_{PBC}$ is unique. One may expect that a ruled surface will appear
on the surface of the convex set $\Theta(\{H_j\})$ consisting of all
the points given by
\begin{equation}\label{eq:VanishingRS}
  \left(\frac{1}{N-2}\tr(H^O_1\rho),\frac{1}{N}\tr(H^O_2\rho),
    \frac{1}{N}\tr(H_3\rho)\right)
\end{equation}
for any quantum state $\rho$, similar as the symmetry-breaking case as
discussed in~\cite{zauner2014symmetry}.

Unfortunately, it is not the case for the SPT order. This is because
that the expectation value of the symmetry-breaking term $H_3/N$ is
mainly contributed from the bulk and the ruled surfaces which result
from the edge states will become invisible. This is essentially the
meaning of `topological' for the SPT orders, in contrast to symmetry
breaking orders. If one instead takes the expectation value of $H_3$,
which is indeed different for the degenerate ground states,\ the set
\begin{equation}
  \left(\frac{1}{N-2}\tr(H^O_1\rho),\frac{1}{N}\tr(H^O_2\rho),
    \tr(H_3\rho)\right),
\end{equation}
which is indeed convex, will be unbounded. Furthermore, if one only
takes the expectation value on the boundary, i.e. to consider
\begin{equation}\label{eq:Nonconvex}
  \left(\frac{1}{N-2}\tr(H^O_1\rho),\frac{1}{N}\tr(H^O_2\rho),
    \tr((X_1 + X_N)\rho)\right),
\end{equation}
then this set is no longer convex (see Appendix~\ref{appen} for more details).

To overcome all these difficulties, we instead add the
symmetry-breaking term on the boundary of the system. This is because
that the degeneracy essentially comes from the edge spins. We
therefore propose to use the following Hamiltonian:

\begin{equation}\label{OBC}
  \begin{split}
    H_{OBC} &= J_1(1+\alpha) H^O_1+J_2(1-\alpha)
    H^O_2 \\
    &- B_x(X_1 + X_N)
  \end{split}
\end{equation}

And for comparison purpose, we also modify the PBC Hamiltonian to be:

\begin{equation}
  \begin{split}
    H_{PBC} & = J_1(1+\alpha) H^P_1+J_2(1-\alpha)
    H^P_2 \\
    & -B_x( X_1 + X_N)
  \end{split}
\end{equation}

For OBC, the convex set $\Theta(\{H_j\})$ can be generated by the
following expectation value with respect to the ground state:
\begin{align}
  & \langle XZX \rangle = \langle H^O_1\rangle/(N-2),\\
  &\langle Z\rangle = \langle H^O_2\rangle/N, \\
  &\langle X\rangle = \langle X_1+X_N\rangle /2.
\end{align}

For PBC, the corresponding quantities are:
\begin{align}
  & \langle XZX \rangle = \langle H^P_1\rangle/N,\\
  &\langle Z\rangle = \langle H^P_2\rangle/N, \\
  &\langle X\rangle = \langle X_1+X_N\rangle /2.
\end{align}

\begin{figure}
  \centering
  \subfloat[]{
    \includegraphics[width=75mm,height=50mm]{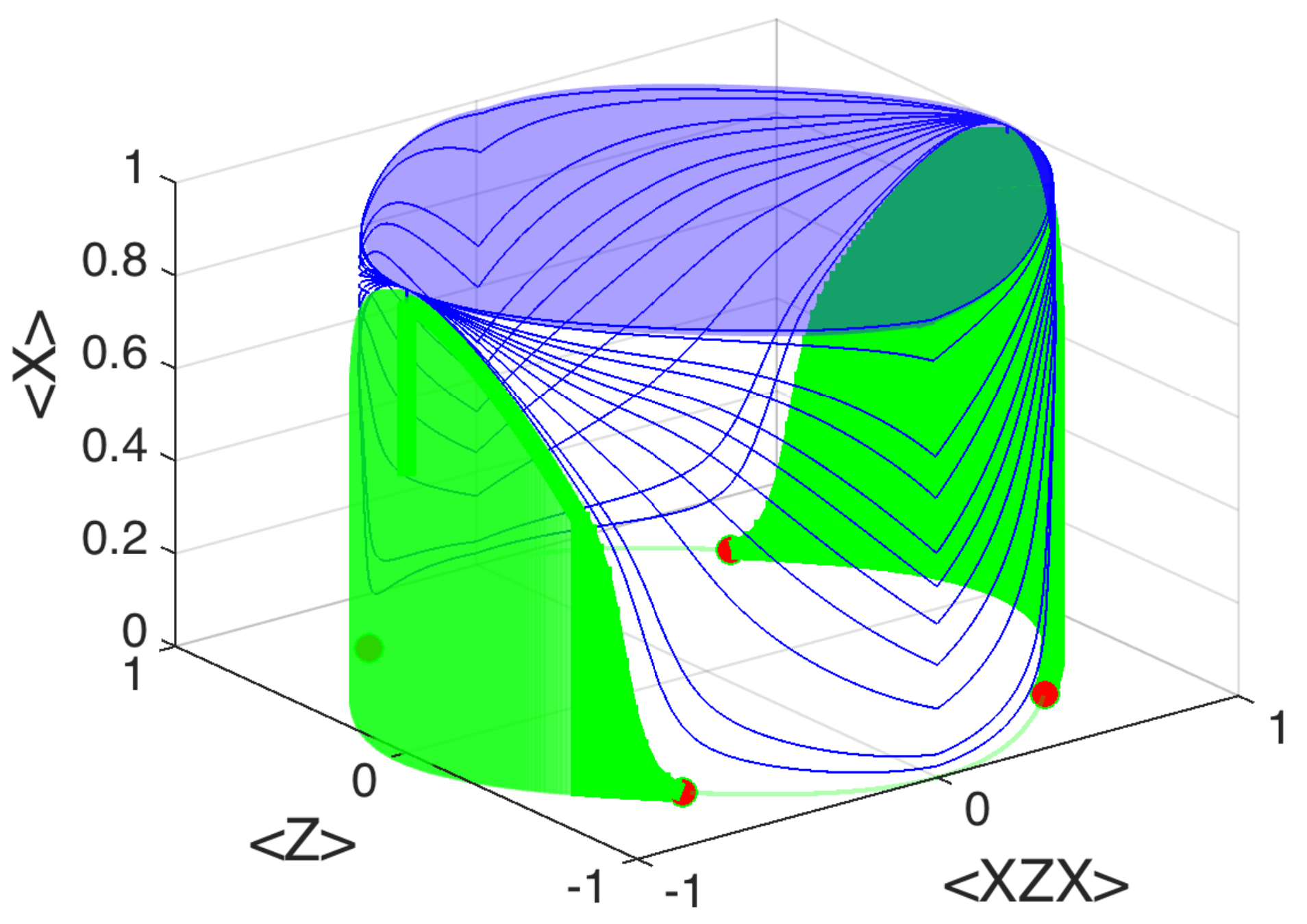}}\\
  \subfloat[]{
    \includegraphics[width=75mm,height=50mm]{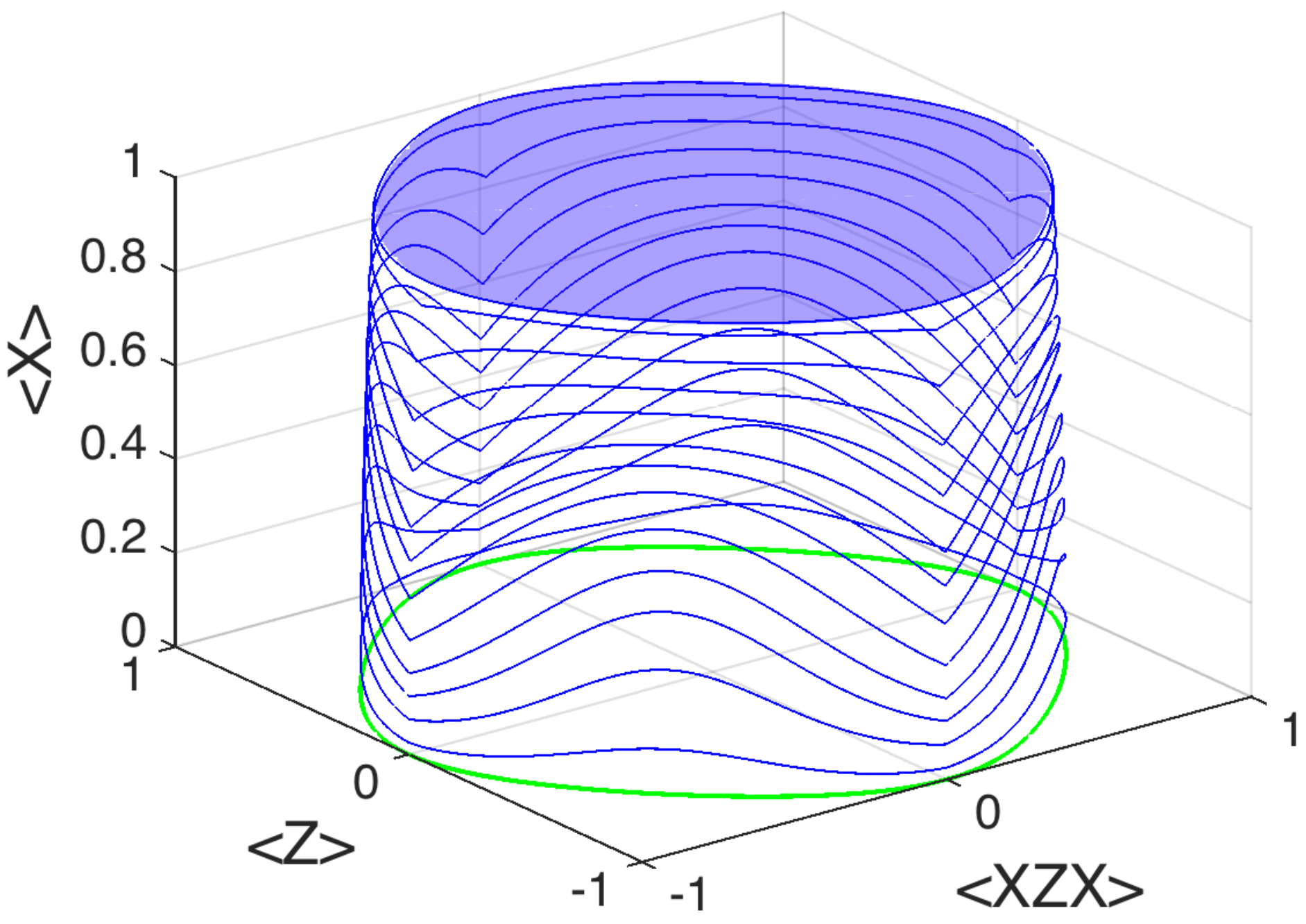}}
  \caption{Schematic figure for the convex set under OBC and PBC. Red
    dots in (a) mark the phase transition point. The green surface in (a) is the ruled surface, which is absent in (b).
    For clarity, only the upper half (i.e. $\langle X\rangle>0$) is shown. For more
    details, see the analysis in Sec.~\ref{sec:algorithm}.}
  \label{schematic_RS}
\end{figure}

We show that there will then be emergence of ruled surfaces on the
boundary of $\Theta(\{H_j\})$ for OBC, and no ruled surfaces for PBC, which is illustrated in
Fig.~\ref{schematic_RS}.

\section{Algorithm and precision}

\label{sec:algorithm}

\begin{figure}
  \centering
  \subfloat[]{
    \includegraphics[width=75mm,height=50mm]{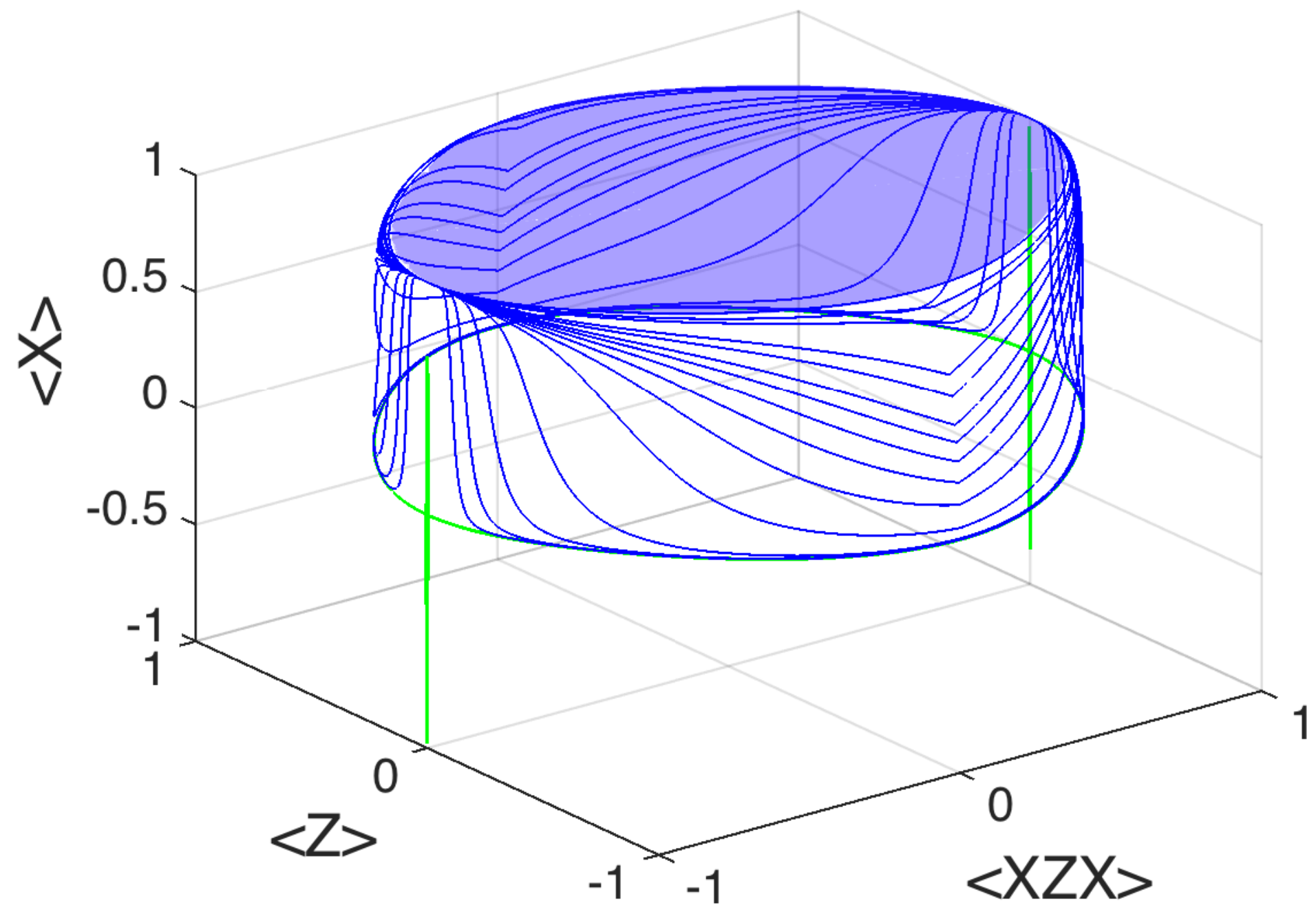}}\\
  \subfloat[]{
    \includegraphics[width=75mm,height=50mm]{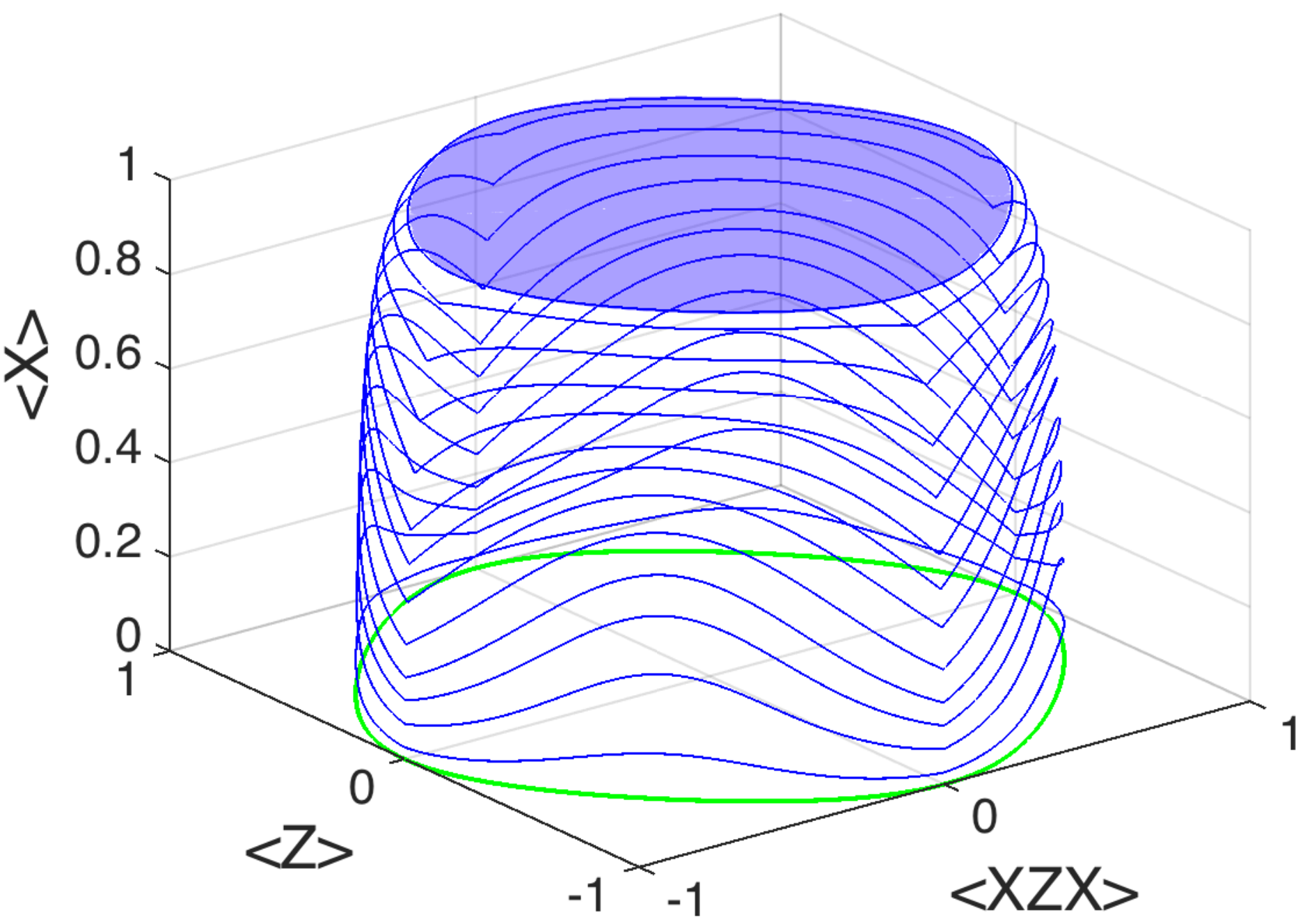}}
  \caption{Convex set for 1D cluster model with system size $N=12$ . (a) represents convex set
    under OBC while (b) is for PBC. For both cases, the $B_x$ field is added only on two
    boundary sites (one for each boundary). The result is obtained
    using ED method. $B_x=0$ is represented by the green line in
    both cases. A dramatic difference between the two cases is that
    a small region reminiscent of ruled surface exists in (a)
    but is absent in (b).}
  \label{ED}
\end{figure}

\begin{figure}
  \centering
  \subfloat[]{
    \includegraphics[width=75mm,height=50mm]{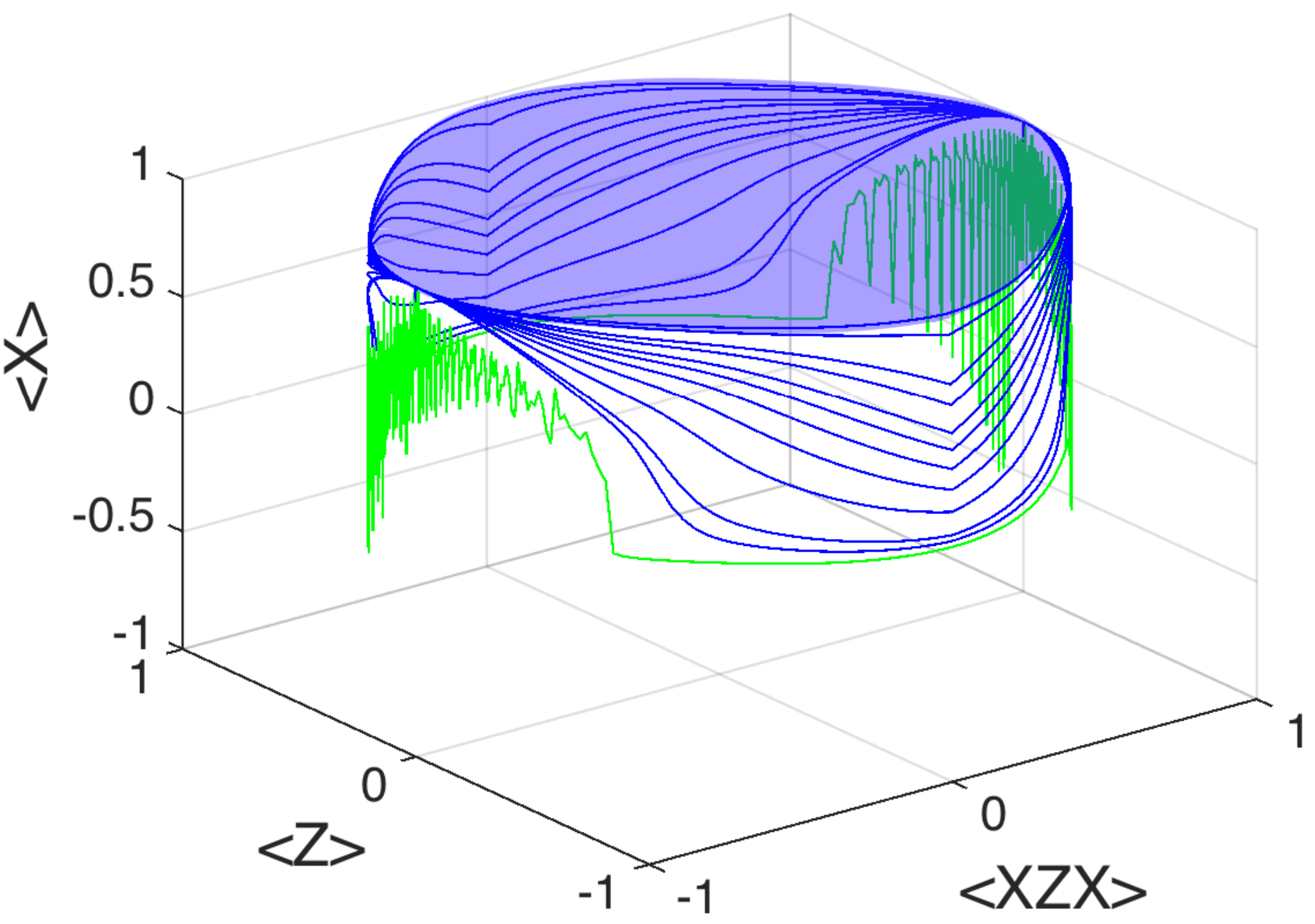}}\\
  \subfloat[]{
    \includegraphics[width=75mm,height=50mm]{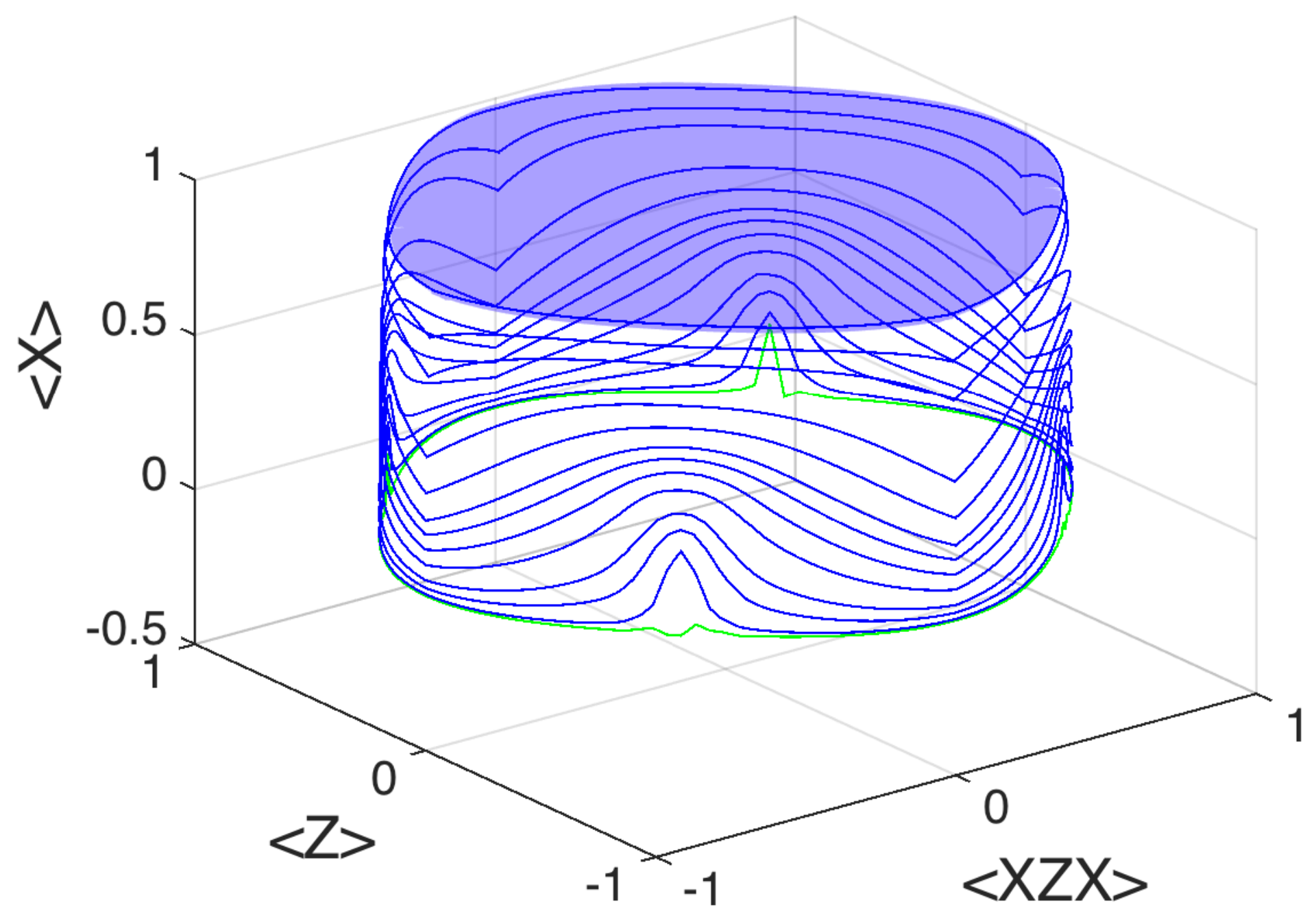}}
  \caption{Convex set for 1D cluster model with system size $N=60$.
    (a) is for OBC while (b) is for PBC. The green vibrational curve is the ruled surface. We use TEBD method
    with internal dimension $D=40$. Due to
    numerical error, we can witness two ruled surface with a large but
    finite system under OBC.}
  \label{MPS}
\end{figure}

We study above models using two different approaches. We first study a
small size system using exact diagonalization (ED) method, where there is no visible error. The numerical result
for $N=12$ is presented in Fig.~\ref{ED}, which already shows the
signal of ruled surface under OBC. To go closer to the thermodynamic limit, we
use matrix product state (MPS) as a variational ansatz and approach
the ground state using Time-Evolving Block Decimation (TEBD)
method~\cite{TEBD_1, TEBD_2, TEBD_3}, whose accuracy is mainly limited by Trotter error and finite dimension of the underlying MPS.
In practical calculations, the ruled surface can be distinguished from a non-ruled one by the oscillating scenario in the convex set which arises due to the ground-state degeneracy, as further discussed in the next paragraph.
 As shown in Fig.~\ref{MPS}, the
green oscillating line indicating the ruled surface indeed only exists under OBC.

\begin{figure}
  \centering
  \subfloat[]{
    \includegraphics[width=75mm,height=55mm]{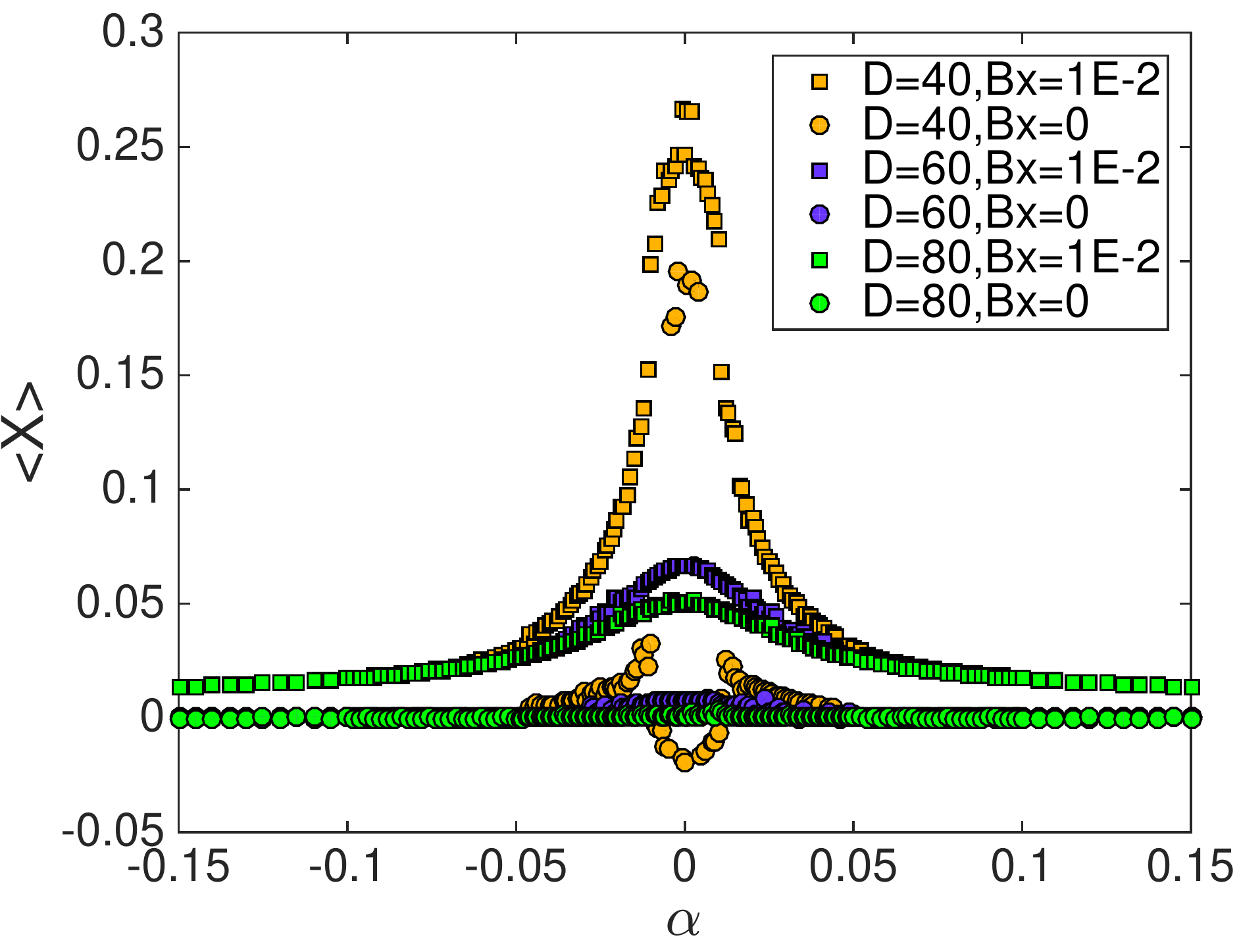}}\\
  \subfloat[]{
    \includegraphics[width=75mm,height=50mm]{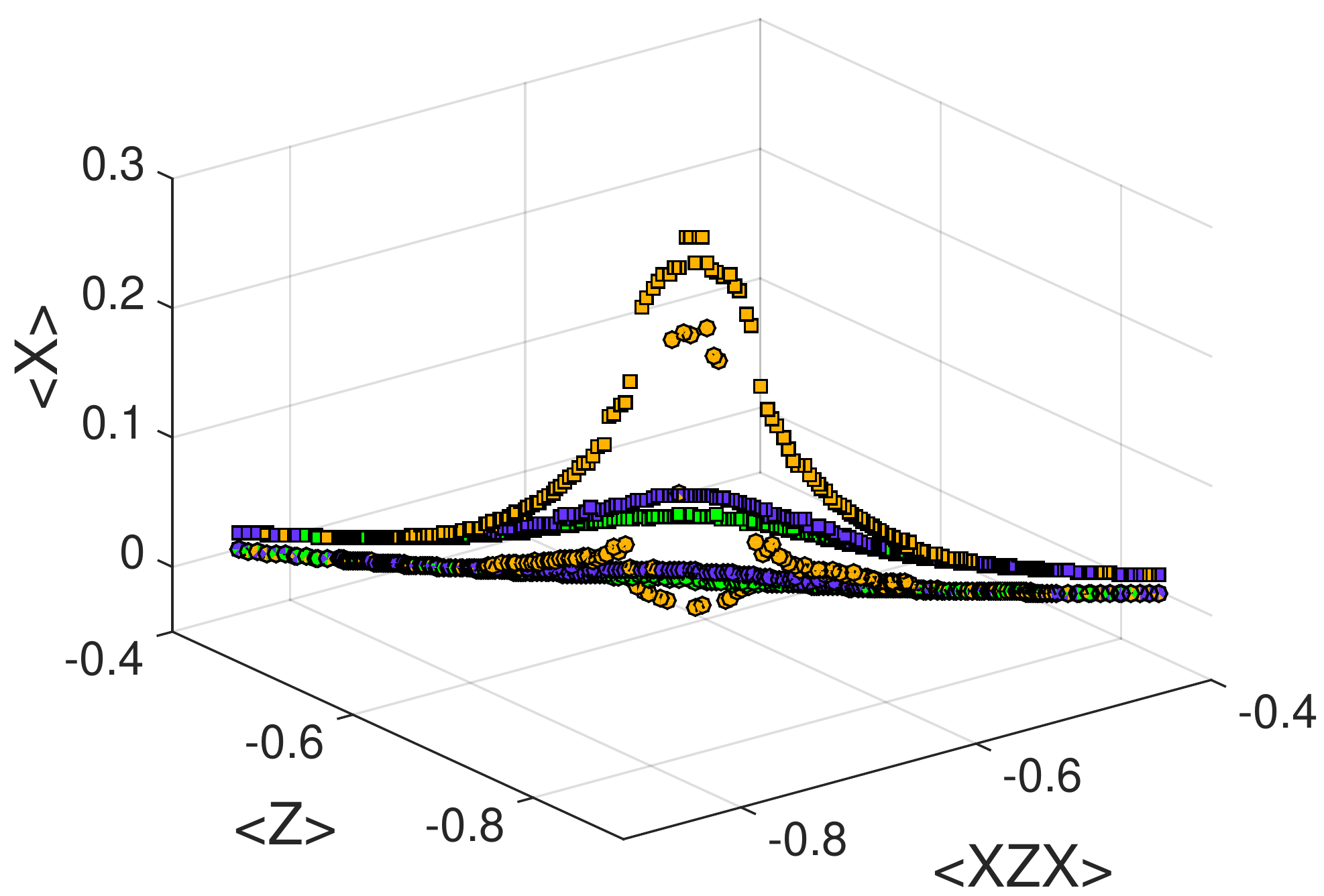}}
  \caption{Finite $D$ scaling for phase transition point under PBC. In
    both (a) and (b), different colors represent different bond
    dimension $D$ where yellow, blue and green is for $D=40,60,80$,
    respectively. Filled square represents $B_x=1\times 10^{-2}$ while
    filled circle represents $B_x = 0$. With increasing $D$, the small
    ruled surface in Fig.~\ref{MPS} (b) near phase transition point shrinks. Eventually, this
    small ruled surface will vanish in the infinite $D$ limit.}
  \label{D_scaling}
\end{figure}

In the thermodynamic limit under OBC, when the
system is in the SPT phase, the ground state space is spanned by 4
degenerate states. For a large finite system, these four states are
nearly degenerate. Thus the state given by TEBD method would be a
superposition of these four states because of the limit of the
numerical accuracy. This explains the vibrational property of the
ruled surface in Fig.~\ref{MPS} (a). As far as we are mainly concerned
with the extent of the ruled surface, it is safe to replace the
original vibrating curve with its upper hull. Similar vibration was
also observed when external magnetic field $B_x$ is small enough, e.g.
$10^{-4}$, which is not shown in the figure.

A thorough investigation of the numeric errors would be both lengthy
and unnecessary. Here we perform a qualitative analysis to show how
such errors interestingly lead to the possibility to obtain the ruled
surface in a large but finite system, while such a surface should only
exist in the thermodynamic limit.

Due the the limit of numerical accuracy, the curve computed for
$B_x=0$ should be more properly understood as the curve corresponding
to $B_x=\epsilon$, where $\epsilon$ is a very small number. Denote by
$f_{N,B_x}$ the curve for a system of size $N$ and the external
magnetic field $B_x$, or the upper hull of it in case of vibration.
Thus our goal is to estimate the difference between $f_{\infty,0}$
(theoretical boundary for the ruled surface in the thermodynamic
limit) and $f_{N,\epsilon}$ (curve observed in a finite system with
numerical errors). Same as finite system, we can take
$f_{\infty,0}\approx f_{\infty,\epsilon}$. Since
$f_{\infty,\epsilon}=\lim_{N\to\infty}f_{N,\epsilon}$, for any
$\delta>0$, there exists $N(\delta)$, such that
$d(f_{\infty,\epsilon}, f_{N,\epsilon})<\delta$ for any
$N\geq N(\delta)$, where $d$ can be taken, for example, to be the
Hausdorff distance between curves. Thus the difference between
$f_{\infty,0}\approx f_{\infty,\epsilon}$ and
$f_{N,0}\approx f_{N,\epsilon}$ can be arbitrarily small for $N$ large
enough. In practice, the convergence is fast such that when $N=60$,
the observed ruled surface precisely represents the ruled surface in
the thermodynamic limit.

For a large finite system under PBC, there seems to be a small ruled
surface at the phase transition point, shown in Fig.~\ref{MPS} (b).
With increasing bond dimension $D$, this small area shrinks and
eventually vanish in the infinite $D$ limit, shown in
Fig.~\ref{D_scaling}. Thus under PBC, there is no ruled surface.

Notice that under both OBC and PBC the upper plane is flat. This is because the normal direction
of the corresponding supporting hyperplane is $(0,0,1)$, which corresponds
to $J_1=J_2=0$ for $H_{OBC}$ and $H_{PBC}$. $H_{OBC}$ and $H_{PBC}$ then both become $-B_x(X_1+X_N)$, which
only acts nontrivially on the boundary, hence are largely degenerate. On the
contrary, each line inside the ruled surface (only under OBC) corresponds to a
(finite) four-fold degeneracy, which is a non-trivial signal of the
SPT order.

\section{Conclusion and discussion}

We study geometry of reduced density matrices for SPT order. Our focus
is on the emergence of ruled surface on the boundary of the convex set
$\Theta(\{H_j\})$. The ground-state degeneracy is a necessary
condition for the existence of those ruled surfaces, yet not
sufficient.

Compared to the ruled surfaces associated with symmetry-breaking order as
discussed in~\cite{zauner2014symmetry}, there is an essential
difference for the SPT order. Since there is no local order parameter
for SPT order, the ruled surface only exists for the open boundary
condition with symmetry-breaking term acting on the boundary. This
term is not a thermodynamic variable. Therefore the emergence of ruled
surface for SPT order is an effect of geometric locality of the
system.

In principle, ruled surface only exists in the thermodynamic limit.
However, we have shown that in practice, finite-size calculation
suffices to reveal this phenomenon, due to inevitable computational
precision uncertainty. This allows us to deal with the calculations
using finite systems.

We hope our discussion leads to further understanding of the geometry
of reduced density matrices, the effect of geometric locality, and SPT
order.

\section*{Acknowledgements}
We thank discussions with Zheng-Cheng Gu, Wenjie Ji and Tian Lan. The work of JYC is supported by the MOST 2013CB922004 of the National Key Basic Research
Program of China, and by NSFC (No. 91121005, No. 91421305, No. 11374176, and No.11404184).
ZJ acknowledges support from NSERC, ARO. ZXL acknowledges the support from NSFC 11204149 and Tsinghua University Initiative Scientific Research Program.
BZ is supported by NSERC and CIFAR.
This research was supported in part by Perimeter Institute for
Theoretical Physics. Research at Perimeter Institute is supported by
the Government of Canada through Industry Canada and by the Province
of Ontario through the Ministry of Economic Development \& Innovation.

\appendix

\section{Vanishing of ruled surface and nonconvex set}
\label{appen}

In this Appendix, we explain in detail why we choose the Hamiltonian
(\ref{OBC}) instead of (\ref{H_OBC}).

\begin{figure}[ht]
  \centering
  \includegraphics[width=75mm,height=50mm]{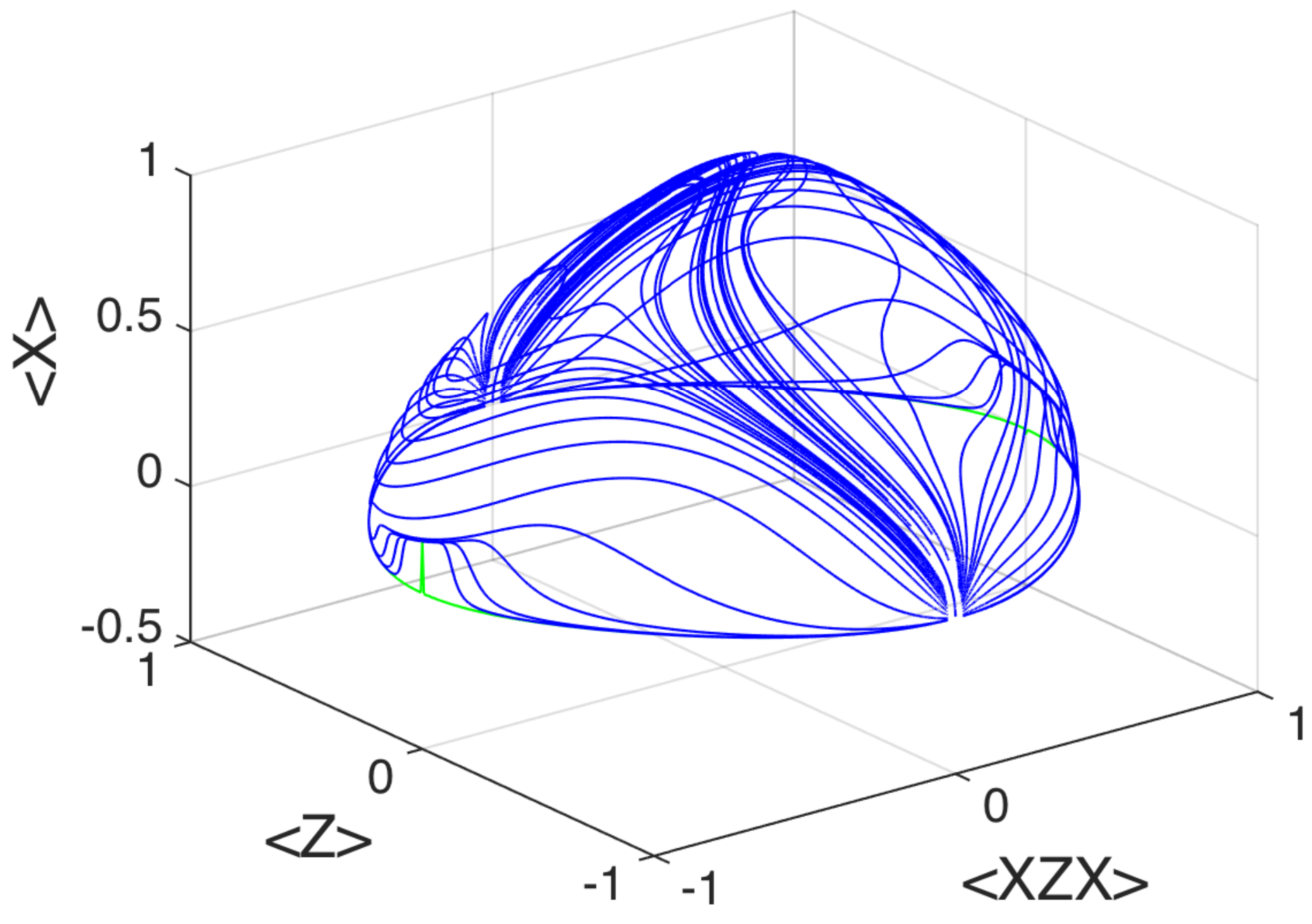}
  \caption{Convex set with a small ruled surface, which is indicated
    by green lines. The result is obtained by ED method
    with $N=12$.
    $\langle XZX \rangle = \frac{1}{N-2}\langle H_1^O \rangle$,
    $\langle Z \rangle = \frac{1}{N} \langle H_2^O \rangle$,
    $\langle X \rangle = \frac{1}{N} \langle H_3\rangle$.}
  \label{vanishRS}
\end{figure}

If we choose the Hamiltonian (\ref{H_OBC}) and plot the convex set
(\ref{eq:VanishingRS}), the ruled surface will vanish in the
thermodynamic limit. The reason is that the degeneracy of the ground
states are owing to the edge states, while the expectation value of
the term ${1\over N}{\rm tr}(H_3\rho)$ mainly comes from the bulk. If
$N\to\infty$, the boundary effect (together with the ruled surface)
will disappear due to the normalization factor ${1\over N}$ (See
Fig.~\ref{vanishRS}).

\begin{figure}[ht]
  \centering
  \includegraphics[width=75mm,height=50mm]{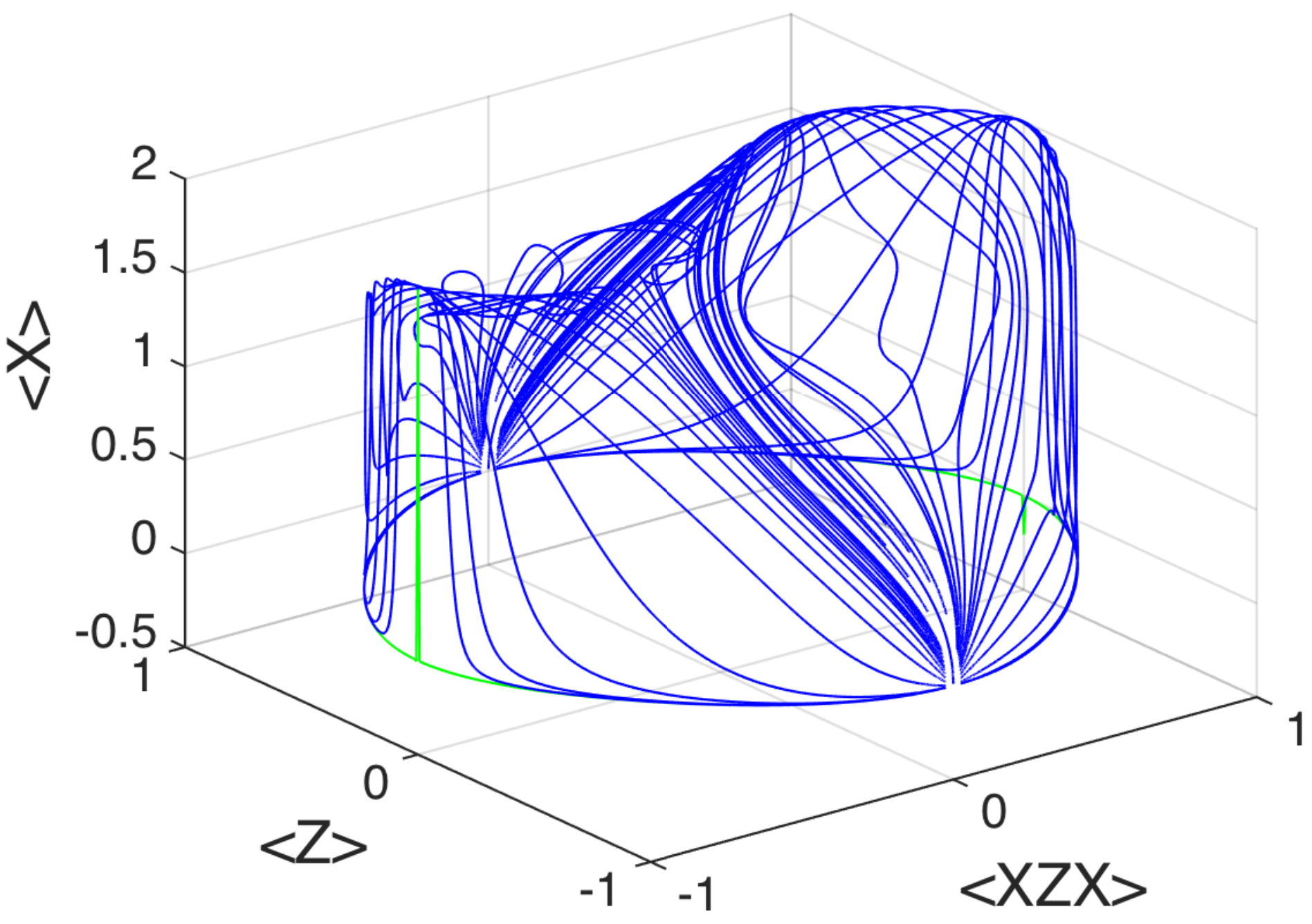}
  \caption{Non-convex set. Result is obtained by ED method
    with $N=12$.
    $\langle XZX \rangle = \frac{1}{N-2}\langle H_1^O \rangle$,
    $\langle Z \rangle = \frac{1}{N} \langle H_2^O \rangle$,
    $\langle X \rangle = \langle X_1 + X_N \rangle$.}
  \label{Nonconvex}
\end{figure}

On the other hand, if we plot the set (\ref{eq:Nonconvex}) to avoid
the vanishing factor $1\over N$, what we obtain is not a convex set
(see Fig.~\ref{Nonconvex}). This is because we cannot use $\{ H_1^O, H_2^O, X_1+X_N\}$
to construct the Hamiltonian (\ref{H_OBC}).

\newpage

\bibliographystyle{unsrt}

\bibliography{RDM-SPT}

\end{document}